\documentclass[aps,twocolumn,showpacs,prl,longbibliography]{revtex4-1}
\usepackage{graphicx}
\usepackage{amsmath}
\usepackage{amsfonts}
\usepackage[table]{xcolor}
\usepackage{array}
\usepackage{floatrow}
\usepackage{pbox}
\usepackage{chngcntr}
\usepackage[colorlinks=true,linkcolor=blue,citecolor=blue,filecolor=cyan,urlcolor=blue,breaklinks=true]{hyperref}

\renewcommand{\vec}[1]{\boldsymbol{#1}}
\newcommand{\mean}[1]{\left\langle #1 \right\rangle}
\newcommand{\Var}{\mathrm{Var}\,}

\newcommand{\peq}{p^\mathrm{eq}}
\newcommand{\fex}{f^\mathrm{ex}}

\newcommand{\dN}{\langle\dot{N}\rangle}
\newcommand{\smean}[1]{\langle #1 \rangle}
\newcommand{\corrN}[1]{\langle\dot{N}(0)\dot{N}(#1)\rangle}

\DeclareMathOperator\erf{erf}

\begin{document}
\title{Classical Pendulum Clocks Break the Thermodynamic Uncertainty Relation}

\author{Patrick Pietzonka}
\affiliation{DAMTP, Centre for Mathematical Sciences, University of
  Cambridge, Wilberforce Road, Cambridge CB3 0WA, United Kingdom}
\affiliation{Max Planck Institute for the Physics of Complex Systems, N\"othnitzer Str. 38, 01187 Dresden, Germany}
\date{\today}

\parskip 1mm

\begin{abstract}
  The thermodynamic uncertainty relation expresses a seemingly
  universal trade-off between the cost for driving an autonomous
  system and precision in any output observable. It has so far been
  proven for discrete systems and for overdamped Brownian motion.  Its
  validity for the more general class of underdamped Brownian motion,
  where inertia is relevant, was conjectured based on numerical
  evidence. We now disprove this conjecture by constructing a
  counterexample. Its design is inspired by a classical pendulum
  clock, which uses an escapement to couple the motion of an
  oscillator to another degree of freedom (a ``hand'') driven by an
  external force. Considering a thermodynamically consistent, discrete
  model for an escapement mechanism, we first show that the
  oscillations of an underdamped harmonic oscillator in thermal
  equilibrium are sufficient to break the thermodynamic uncertainty
  relation. We then show that this is also the case in simulations of
  a fully continuous underdamped system with a potential landscape
  that mimics an escaped pendulum.
\end{abstract}

\maketitle

\paragraph{Introduction.}Being able to tell the time precisely, regardless of
astronomical observation, has ample importance for virtually all of human
civilisation. Ancient water clocks or hour glasses relied on the steady flow of
matter. However, the precision of such devices was strongly limited, because
of inhomogeneities or unpredictable external influences. A revolution in the
history of timekeeping was the invention of the escapement. This mechanism
uses coherent oscillations of a physical system to regulate the forward motion
of a cog loaded with a weight and connected to a hand that displays
time. Galileo realised that a swinging pendulum can provide such oscillations,
since its period is (for small angles) independent of its amplitude. This
inspired Huygens' invention of the pendulum clock in 1656, setting standards
in precision for centuries to come~\cite{marr48}. This Letter shows that this
well-established principle even allows for precision beyond the thermodynamic
limits that have so far been believed to apply to classical systems.

The performance of a clock can be quantified by its precision and its turnover
of energy. These quantities take center stage in the thermodynamic uncertainty
relation (TUR)~\cite{horo19}, formulated originally for biomolecular
systems~\cite{bara15}. It describes a trade-off between the overall cost
for driving a system and the precision observed in any output current.

More specifically, we consider a Markovian system in a steady state
producing an integrated current~$Y(t)$ (e.g., the accumulated angle of
a clock hand). The energetic cost of driving is quantified by the
entropy production rate $\sigma$. It corresponds (in the absence of
chemical changes) to the heat dissipated into a surrounding heat bath,
divided by its constant temperature~$T$. This heat needs to be equal
to the energy expended on the system's driving. The TUR states that
\begin{equation}
  \label{eq:TUR}
  \frac{\Var{Y(t)}}{\mean{Y(t)}^2}t\sigma\geq 2,
\end{equation}
where we set Boltzmann's constant $k_B=1$ and define
$\Var{Y(t)}\equiv\mean{Y(t)^2}-\mean{Y(t)}^2$, with averages $\mean{\ldots}$
taken in the steady state. This relation was first proven for the limit of
large times $t$ \cite{ging16}, and later generalised to finite
times~\cite{piet17,horo17}.

The TUR rests on the premise of local detailed balance, relating the
log-ratio of forward and backward transition rates between discrete
states of a system to the entropy produced in a
transition~\cite{seif18}. Brownian diffusion fits into this framework
if it can be described as \textit{overdamped}, meaning that momentum
variables are assumed to relax instantly to a local equilibrium. Here, the TUR is recovered either
through a fine discretisation of the state space or directly from a
Langevin description~\cite{dech18a,dech18}.

An underdamped description of Brownian dynamics explicitly retains the inertia
that is present in every classical dynamical system. Yet, this more general
dynamics lies beyond the original framework of the TUR. There, transitions in
a finely discretised phase space appear irreversible without the simultaneous
reversal of momenta, which leads to a formally divergent entropy production,
despite the actual entropy production being finite~\cite{fisc18}. Bounds on
the precision of irreversible currents in underdamped systems have been
derived~\cite{fisc18,van19,lee19,vo20,lee21}, however, they are weaker than
the TUR or require additional information about the system (beyond the entropy
production rate). Violations of the TUR in its original form have been
observed where an external magnetic field breaks time-reversal
symmetry~\cite{bran18,chun19}. It is also known that ballistic motion on short
timescales spoils the finite-time variant of the TUR. Yet, numerical evidence
suggested that the TUR would hold for large times~\cite{fisc20}, in line with the
intuition that sufficient driving is needed to overcome the time-reversal
symmetry of thermal equilibrium.

In this Letter, we provide a disproof of the TUR for underdamped
dynamics. As it turns out, the principle of an escapement, originally
conceived to ward off environmental perturbations, is effective even
if these are of purely thermal origin. We develop a minimal
thermodynamically consistent model for a pendulum clock, which yields
at given energetic cost high precision beyond the limits of the TUR.

Once wound, pendulum clocks operate autonomously. Hence they are
different from periodically driven systems, for which refined
TUR's have been
derived~\cite{bara16,proe17,koyu18,bara18,koyu20}. Taken for
themselves, periodic systems do not obey the original TUR. It nonetheless
holds on a global scale taking into account the infinite cost
required to generate a deterministic protocol in a Markovian
framework \cite{bara17}. We take this idea further, describing an
escapement as a way to couple a discrete system to an oscillating
system that is not perfectly precise, but comes at small (or even zero)
energetic cost.

\begin{figure}
  \centering
  \includegraphics[width=\textwidth]{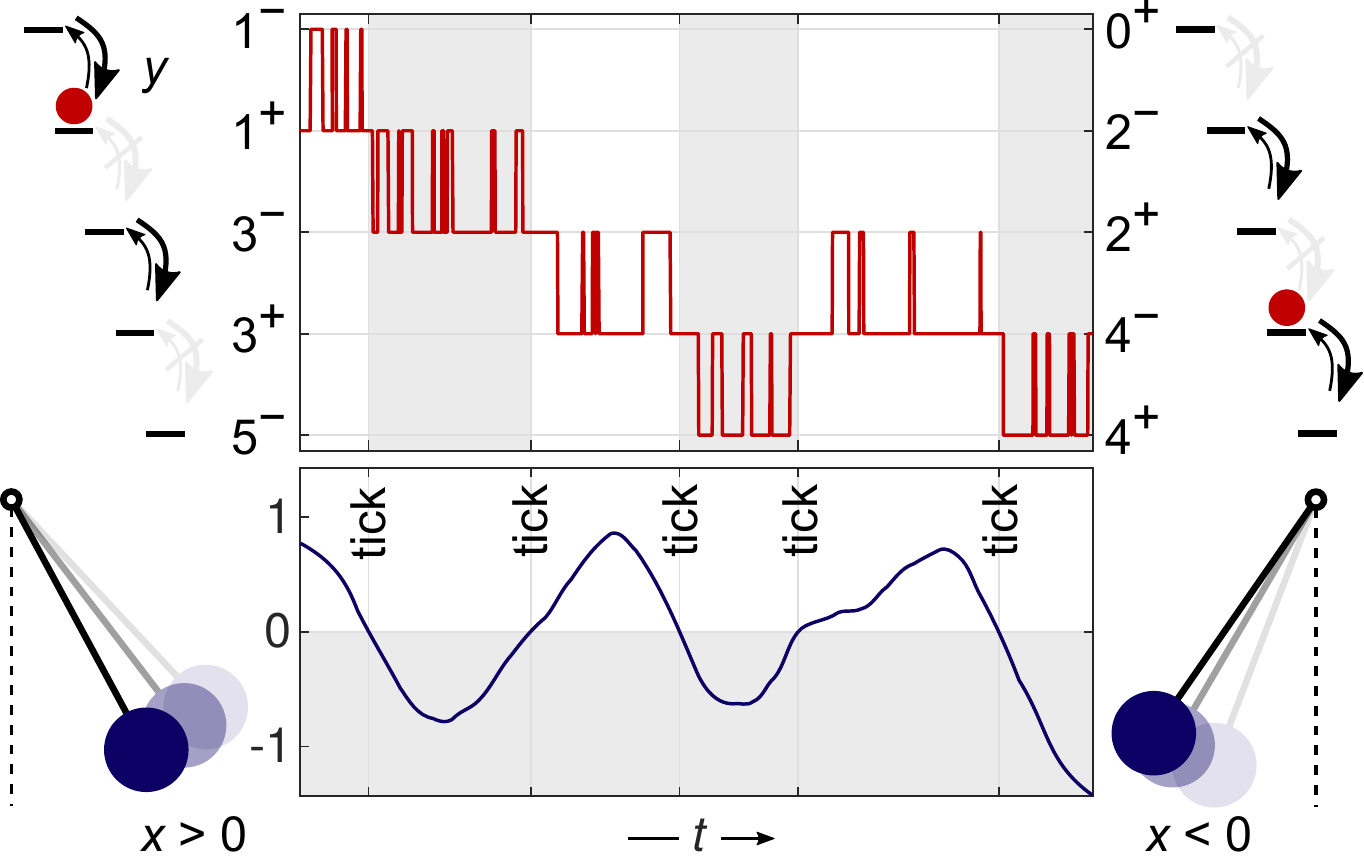}%
  \caption{Minimal model for a pendulum clock subject to thermal
    noise. An oscillating degree of freedom $x$ (bottom), is coupled
    to a discrete counter degree of freedom (top), where $y$ labels
    pairs of adjacent states $y^-$ and $y^+$. A typical trajectory is
    shown in the middle, with snapshots of the configuration at its
    beginning (left) and end (right). Transitions of the counter are
    generally biased downward. When $x>0$, transitions between even
    pairs of states $y^\pm$ are strongly suppressed and transitions
    between odd ones are very fast, and vice versa for $x<0$. This
    way, transitions from one pair of states $y$ to a neighbouring one
    are possible only upon a ``tick,'' when the pendulum crosses
    through $x=0$. }
  \label{fig:escapement}
\end{figure}

\paragraph{General setting.} We consider a system consisting of two
subsystems, as shown in Fig.~\ref{fig:escapement}. The first, which we
call an ``oscillator,'' could be an arbitrary physical
system described by a state $x(t)$ (in our main example and
Fig.~\ref{fig:escapement} we choose this to be a pendulum subject to thermal noise). The other
subsystem, which we call a ``counter,'' is a one-dimensional degree of
freedom on an infinite discrete lattice. For notational convenience we label
pairs of adjacent states by integers $y$, denoting the upper state as
$y^-$ and the lower state as $y^+$, and identifying $y^+$ and
$(y+1)^-$ as the same states of the counter. In the picture
of a pendulum clock, the discrete state space of the counter corresponds to
orientations of the seconds hand, mapped to the infinite number line
by keeping account of full revolutions around the clock face.

Jumps between any $y^-$ and the corresponding $y^+$ occur continuously in time
and are biased toward the latter by a nonconservative force (e.g., provided
by a weight on a cord). The work delivered by this force in a forward jump
divided by the temperature $T$ defines the affinity $A$, fixing the log-ratio
of the rates for forward and backward transitions through the local detailed
balance condition
\begin{equation}
  \label{eq:ldb}
  \ln\frac{k(y^-\to y^+)}{k(y^+\to y^-)}=A.
\end{equation}
Since the states $y^-$ and $y^+$ have the same internal energy, the first law
requires the work to be dissipated as heat, increasing
the entropy in the environment by $A$. Conversely, a backward step decreases this entropy by $A$.

The escapement mechanism is realised by exploiting the freedom of
choice of a common prefactor to both rates, which can be made
dependent on the state $x$ of the oscillator. We model that
the escapement can be in either of two states $i(x)\in\{0,1\}$. If $x$
is the angular displacement of a pendulum, an obvious choice is $i(x)=1$ for
$x\geq 0$ and $i(x)=0$ for $x< 0$. For $i(x)=1$, we set the transition rates
$k(y^\mp\to y^\pm)=k^\pm$ for $y$ odd and $k(y^\mp\to
y^\pm)=\varepsilon k^\pm$ for $y$ even. Vice versa, for $i(x)=0$, we set
$k(y^\mp\to y^\pm)=\varepsilon k^\pm$ for $y$ odd and $k(y^\mp\to
y^\pm)=k^\pm$ for $y$ even. The rates $k^+$ and $k^-=k^+\exp(-A)$ are both
chosen to be much larger than the inverse of the fastest relevant
timescale of the oscillator. In contrast, the factor
$\varepsilon>0$ is chosen sufficiently small such that the rates
$\varepsilon k^\pm$ are much smaller than the inverse of the slowest
timescale of the oscillator.
This choice of rates ensures that, after a change of the state $i(x)$, the counter
is effectively constrained to a single pair of states $y^\pm$, between which it
equilibrates quickly. It can then be found in either of the states at
the conditional probability
\begin{equation}
  \label{eq:condprob}
  p_\pm=k^\pm/(k^++k^-)=\exp(\pm A/2)/[2\cosh(A/2)].
\end{equation}
Next time the state $i(x)$ changes, the counter will have ended up in
either of the states $y^\pm$, whose link then gets effectively
broken. Depending on this outcome, the counter then proceeds to
fluctuate either between states $(y-1)^\pm$ or $(y+1)^\pm$.

We see that upon every change of $i(x)$, called a ``tick'' in the
following, the variable $y$, labeling the pair of states between which
the counter currently fluctuates, performs a single step of an
asymmetric random walk. In this random walk, the number of ticks
$N(t)$ up to time $t$ plays the role of a discrete time. The steps
taken by $y$ upon subsequent ticks are independent and identically
distributed, such that (given $y=0$ at time $t=0$) the central limit
theorem yields the conditional probability $p(y|N)$ as a Gaussian with
mean and variance
\begin{align}
  NJ_{y|N}&\equiv N(p^+-p^-)=N\tanh(A/2),\\
  2ND_{y|N}&\equiv N[p^++p^--(p^+-p^-)^2]=N/\cosh^2(A/2).
\end{align}

While the dynamics of the counter depends strongly on the dynamics of the
oscillator, there is no feedback in the other direction. This is possible in
our model because a change of $i(x)$ leaves the energy levels of the counter
intact, such that no work is transferred between the two subsystems. In
practice, where the discrete dynamics of the counter is derived from the
diffusion in a corrugated potential, the reduction of rates by the
factor $\varepsilon$ requires the insertion of some potential barrier. This
can be achieved at the expense of a vanishingly small amount of work, by
finely tuning the width and height of the barrier~\cite{bara16}.

With this insight, we see that the process $N(t)$ counting the number
of ticks generated by the oscillator up to time $t$ is \textit{a
  priori} independent of the counter $y$. Given the distribution
$p(N,t)$, the distribution of the state of the counter at time $t$
follows as
\begin{equation}
  p(y,t)=\sum_Np(y|N)p(N,t).
\end{equation}
We now assume that the dynamics of the oscillator is such that $N(t)$
satisfies a central limit theorem with mean $\dN t$ and
variance $2D_Nt$, as will be the case for the harmonic oscillator
considered below. Then, the distribution $p(y,t)$ is Gaussian as well
with mean and variance
\begin{align}
  \mean{y(t)}&=\dN J_{y|N}t,
  \label{eq:meany}\\
  \Var{y(t)}&=2(D_{y|N}\dN+D_NJ_{y|N}^2)t.
    \label{eq:vary}
\end{align}
The entropy production rate of the counter is given by
\begin{equation}
  \sigma_\textup{ctr}=A\dN J_{y|N},
    \label{eq:sctr}
\end{equation}
which is the affinity of a single step multiplied by the net rate of
forward steps. The entropy production rate for the total system
$\sigma=\sigma_\textup{ctr}+\sigma_\textup{osc}$ follows by adding the
entropy production of the oscillator $\sigma_\textup{osc}$.

Given the three relevant quantities for the oscillator,
$\sigma_\textup{osc}$, $\dN$, and $D_N$, we can now check whether a
TUR of the form~\eqref{eq:TUR} holds for the observable $y$ for all
values of the affinity $A$. If it does not, then the TUR cannot be
valid for the type of dynamics underlying the oscillator.

\begin{figure}
  \centering
  \includegraphics[width=\textwidth]{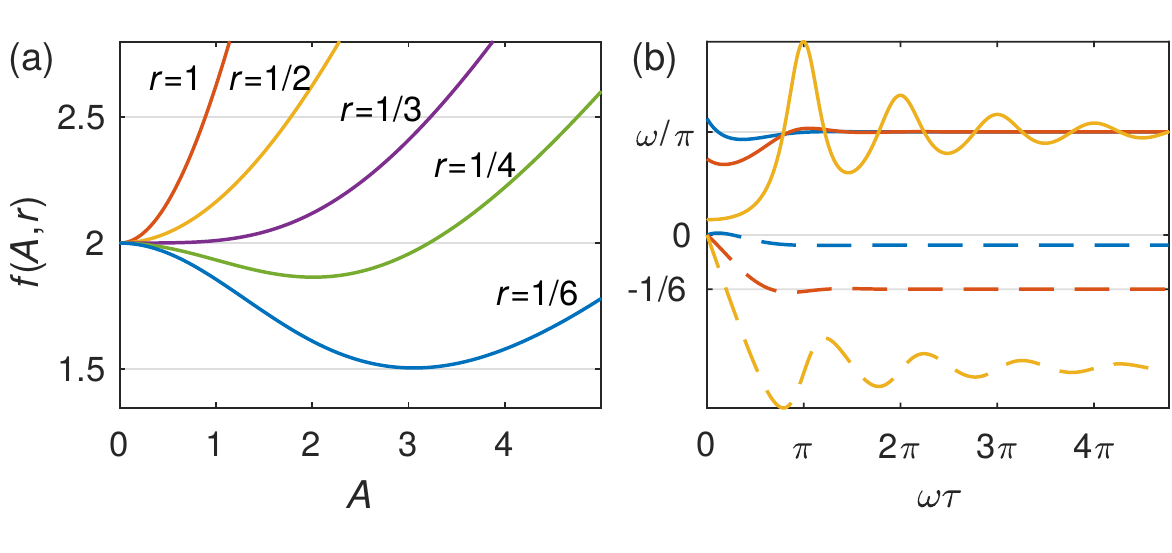}
  \caption{(a) Plot of the function $f(r,A)$ of Eq.~\eqref{eq:fdef}, for
    selected values of the parameter $r$. The value $r=1/3$ is a
    critical one, below which a minimum below $1$ occurs at nonzero
    $A$. (b) Correlation function $\corrN{t}/\dN$ (solid) and its integral $\int_{0^+}^t d\tau[\corrN{\tau}/\dN-\dN]$ (dashed). The parameter $\gamma/(m\omega)$ has
  values $1.5$ (blue), $0.981$ (red), and $0.2$ (yellow).}
  \label{fig:correlation}
\end{figure}

In particular, if the oscillator system is in thermal equilibrium, the
only entropy production is that of the counter, and the
product of relative uncertainty and entropy production becomes
\begin{equation}
  \label{eq:uncprod}
 \lim_{t\to\infty}\frac{\Var{y(t)}}{\mean{y(t)}^2}t\sigma=f(A,D_N/\dN)
\end{equation}
with the function
\begin{equation}
  \label{eq:fdef}
  f(A,r)=2A[1/\sinh(A)+r\tanh(A/2)],
\end{equation}
shown in Fig.~\ref{fig:correlation}(a). For $r\geq 1/3$, it has the global minimum
$2$, attained for $A\to 0$. Hence, for $D_N/\dN\geq 1/3$,
an inequality in the form of the TUR holds. However, for
$D_N/\dN<1/3$ the relation is broken for sufficiently small
values of $A$.

\paragraph{Harmonic oscillator in equilibrium.} We now specify the
oscillator system to be a pendulum, modeled as an underdamped harmonic
oscillator in a heat bath at the same temperature $T$ as the heat bath
of the counter. The position $x$ and velocity $v$ obey the Langevin
equation
\begin{align}
  \dot x=v,\qquad
  m\dot v=-m\omega^2x-\gamma v+\xi(t),
  \label{eq:langevin}
\end{align}
where the dot denotes a time derivative, $m$ is the mass, $\omega$ the
undamped angular frequency, and $\gamma$ the damping coefficient. The
Gaussian white noise $\xi(t)$ has zero mean and correlations
$\mean{\xi(t)\xi(t')}=2\gamma T\delta(t-t')$. The equilibrium state is
the Gaussian $\peq=\exp[-E(x,v)/T]/Z$ with the energy
$E(x,v)=m(\omega^2x^2+v^2)/2$ and normalisation~$Z$.

Partitioning the state space into $i(x)=1$ for $x\geq0$ and $i(x)=0$
for $x<0$, ticks occur whenever $x$ crosses through zero. The total
number of ticks up to time $t$ is
\begin{equation}
  N(t)=\int_0^td\tau\,|v(\tau)|\,\delta[x(\tau)],
\end{equation}
where the factor $|v|$ ensures that every
crossing of $x=0$ at speed $v$ increments $N(t)$ by one.
The average rate of ticks follows readily from the equilibrium
distribution as
\begin{equation}
  \dN=\int dx\int dv\,\peq(x,v)\,|v|\,\delta(x)=\omega/\pi.
\end{equation}

The dispersion of ticks is calculated as~\cite{seif10}
\begin{equation}
  \label{eq:GreenKubo}
  D_N=\frac{1}{2}\dN+\int_{0^+}^\infty d\tau\,[\corrN{\tau}-\dN^2].
\end{equation}
The lower limit $0^+$ indicates that the trivial self-correlation of
every tick at $\tau=0$ is excluded, it produces the first term. The
correlation function $\corrN{\tau}/\dN$ [shown
in Fig.~\ref{fig:correlation}(b)] can be interpreted as the probability
density of a tick occurring at time $\tau$ given a tick
at time $0$. We calculate it analytically from the Gaussian
propagator~\footnote{See Supplemental Material at [URL will be inserted by
  publisher] for the calculation of the correlation function of ticks and a
  derivation of the timescale seapration between counter and oscillator.} and evaluate the integral in
Eq.~\eqref{eq:GreenKubo} numerically. The mean and variance of the counter
variable $y(t)$ can then be calculated using Eqs.~(\ref{eq:meany} and
\ref{eq:vary}), provided the timescale separation $k^\pm\gg \omega/\pi$ and
$k^\pm\gg\gamma/m$~\cite{Note1}. For small
damping $\gamma$, the correlation function exhibits oscillations with
maxima at subsequent ticks, which are initially sharply peaked and
then become broader. The time in between these peaks can have a
sufficient negative contribution, so that the overall integral becomes less than
$-\dN/6$, yielding $D_N/\dN<1/3$. This is the case for the
damping below a certain critical value, determined numerically as
$\gamma/(m\omega)\simeq 0.981$. Remarkably, this critical value presents
still a fairly strong damping, with just one coherent oscillation
discernible in Fig.~\ref{fig:correlation}b. For any damping weaker
than that, the TUR is violated for matching affinity $A$.

In the limit of vanishing damping, $\gamma\to 0$, the sequence of
ticks becomes deterministic (regardless of the energy, which is sampled
initially from $\peq$). The counter system then behaves as a
discrete-time Markov process, for which the possibility of violating
the TUR~\eqref{eq:TUR} is well known~\cite{shir17,proe17}. Yet, we show
here that such a discrete-time process can be realised as a limiting
case of a continuous one, without additional entropic cost. In
accordance with the discrete-time TURs of Refs.~\cite{proe17} and~\cite{liu20}, our model
allows for a vanishing uncertainty product~\eqref{eq:uncprod} for
$D_N/\dN\to 0$ and $A\to\infty$. The latter entails either divergent
entropy production or vanishing speed $\langle\dot y\rangle$. For
clocks that require the hand to move forward at a nonvanishing speed, a
recent study shows that precision does indeed come at a minimal
energetic cost~\cite{pear21}.

\paragraph{Continuous model.} So far, we have shown that the TUR does not hold
for systems consisting of a discrete and an underdamped continuous degree of
freedom. We now show numerically that the TUR can also be broken with two
continuous degrees of freedom. Moreover, we consider an escapement that, like
in actual pendulum clocks, provides a feedback on the oscillator to sustain
amplitudes beyond those of equilibrium oscillations.

We use an underdamped Langevin equation of the form
\begin{equation}
  m\ddot{\vec r}=-\nabla
  V(\vec{r})-\gamma\dot{\vec{r}}+\fex\vec{e}_y+\vec{\xi}(t)
  \label{eq:langevin2}
\end{equation}
for the configuration $\vec{r}=(x,y)^T$, with the mass $m$, damping
$\gamma$, a driving force $\fex$ acting in the $y$ direction, and a noise
term $\vec{\xi}(t)$ with two independent components with the same
properties as for the harmonic oscillator above. The potential is
harmonic in $x$, with an additional coupling term,
$V(\vec{r})=m\omega^2x^2/2+V_\textup{c}(\vec{r})$.

\begin{figure}
  \centering
  \includegraphics[width=\textwidth]{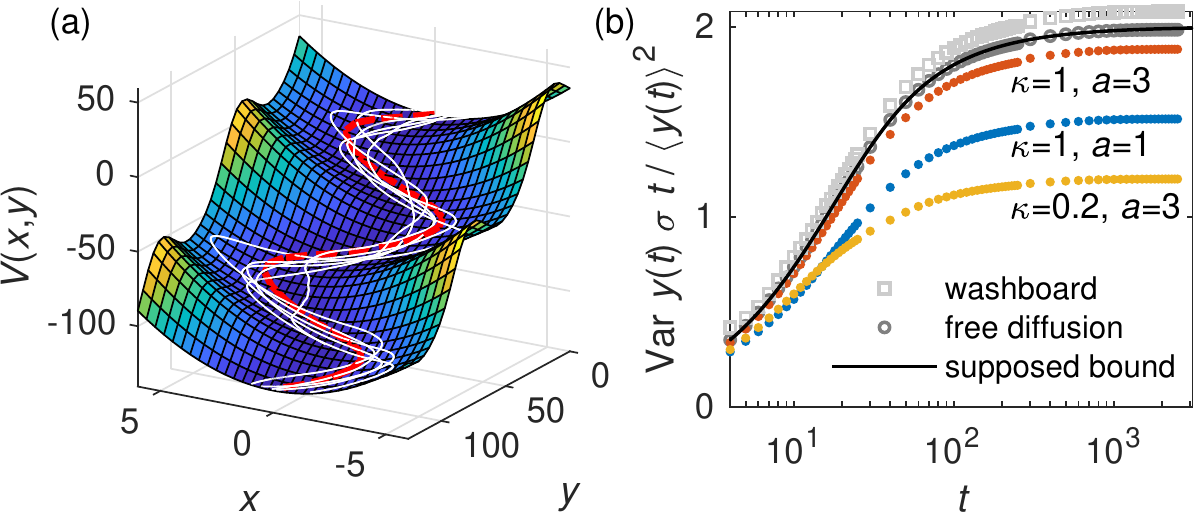}
  \caption{(a) Potential landscape $V(x,y)$ serving as a continuous
    model for an escapement. The external force $\fex=1$ is indicated
    as a tilting in the $y$ direction. The ideal curve $\hat x(y)$ is
    shown as red dashed, with sample trajectories in
    white. Parameters: $\kappa=1$, $a=3$, $\omega=1$, $\gamma=0.1$,
    $T=1$, $m=1$. (b) Uncertainty product as a function of the length
    $t$ of the time window. The bound conjectured in
    Ref.~\cite{fisc20} (solid line) corresponds to free diffusion in
    $y$ ($\kappa=0$, dark grey). This bound holds for the washboard
    potential
    $V_c(\vec{r})=(\fex \gamma/\omega) \sin(\omega y\gamma/\fex)$
    (light grey). For the escapement potential the bound is broken. In
    particular, the long-time limit is below the value $2$ relevant
    for the TUR~\eqref{eq:TUR} for all three combinations of the
    parameters $\kappa, a$ shown (red, blue, yellow). All other
    parameters as in (a).}
  \label{fig:hamiltonian}
\end{figure}

We choose the coupling term such that it reinforces the harmonic
motion of an undamped oscillator of frequency $\omega$ and a certain
amplitude $a$ in the $x$ direction while moving steadily at terminal
velocity $\fex/\gamma$ in the $y$ direction. This ideal motion traces out
the curve $\hat x(y)=a\sin(\omega y\gamma/\fex)$, and we choose the
coupling potential such that this curve is favoured, setting
$V_\textup{c}(\vec{r})=\kappa [x-\hat x(y)]^2/2$, with some stiffness
$\kappa$. Fig.~\ref{fig:hamiltonian}(a) shows this potential and the
ideal curve. For suitably chosen $\kappa$ and $a$, sample trajectories
follow the ideal curve closely.  The potential landscape acts as an
escapement, providing potential barriers that impede the motion in
the $y$ direction when it is too fast, and accelerates it when it is too
slow.

We simulate the steady state of the system by numerically
integrating the Langevin equation~\eqref{eq:langevin2}, using the
method of Ref.~\cite{groe13}. The variance of $y(t)$ is calculated for
samples of time windows of length $t$ taken from a long
trajectory. The current $J_y=\mean{y(t)}/t$ (for any $t$) is evaluated
as the average speed over the whole trajectory, and the entropy
production rate follows as $\sigma=\fex J_y/T$.

As a result, the escapement mechanism suppresses fluctuations in
the $y$ direction, compared to the fluctuations of an underdamped particle
diffusing freely, see Fig.~\ref{fig:hamiltonian}(b). In Ref.~\cite{fisc20}, it
had been conjectured that free diffusion sets a lower bound on the uncertainty
product [the left hand side of the TUR~\eqref{eq:TUR}] for underdamped
dynamics on finite timescales. While this supposed bound holds true for most
generic potential landscapes (and in particular for uncoupled dynamics, e.g.,
a washboard potential in the $y$ direction independent of $x$), it is broken for
our design of an escapement. Violations occur for any $t>0$, and in particular
in the long-time limit, over a robust range of the parameters $\kappa$ and
$a$.

\paragraph{Outlook.} We have used a simple design of an escapement coupled to
a pendulum to construct a counterexample to the TUR for underdamped
dynamics. The considerations that have led to
Eqs.~\eqref{eq:meany}-\eqref{eq:sctr} are specific for the model of
the escapement, but completely general about the oscillator producing
the ticks. The application to other physical systems may also be
fruitful. For instance, for quantum systems exhibiting coherent
oscillations, a general TUR of the form \eqref{eq:TUR} could also be
ruled out, in line with previous
observations~\cite{bran18,agar18,ptas18}. An atomic clock could hence
yield precision beyond the limitations of the TUR as well.

Likewise, a
thermodynamically consistent analysis of resistor-inductor-capacitor (RLC) circuits \cite{frei20}
could reveal coherent oscillations similar to those of the underdamped
oscillator, showing that the TUR is broken not only for constant,
external magnetic fields but also for fluctuating magnetic fields
generated by the system itself. Steady state thermoelectric devices exploiting this
fact could evade the trade-off between power, efficiency, and
constancy that follows from the TUR~\cite{piet18}, similar to cyclically
driven heat engines~\cite{shir16,holu18}.

Future research, systematically comparing different designs of
escapement mechanisms and oscillator systems, may reveal ultimate thermodynamic
bounds on the precision of autonomous clocks and complement
thermodynamic uncertainty relations for underdamped dynamics.

\begin{acknowledgments}
  \paragraph{Acknowledgments.} I
  thank Michael E. Cates for discussions and comments on the
  manuscript. I also thank Corpus Christi College, Cambridge for support
  through a nonstipendiary Junior Research Fellowship and inspiration
  about pendulum clocks. This work was funded by the European Research
  Council under the EU's Horizon 2020 Program, Grant No.~740269. 
\end{acknowledgments}

\bibliography{refs}

\clearpage

\onecolumngrid
\counterwithout{equation}{section}
\setcounter{section}{0}
\setcounter{equation}{0}
\renewcommand\theequation{S.\arabic{equation}}

\section*{Supplemental Material: Classical pendulum clocks break the thermodynamic uncertainty relation}

\subsection{Correlations of the underdamped harmonic oscillator in equilibrium}
\label{sec:ho}

First, we make the system dimensionless by scaling time as $\tilde
t\equiv \omega t$, space as $\tilde x\equiv x\sqrt{m/T}/\omega$ and
velocity as $\tilde v\equiv v \sqrt{m/T}/\omega^2$. The Langevin
equation~(12) of the main text then becomes
\begin{align}
  \partial_{\tilde t}\tilde x&=\tilde v,\nonumber\\
  \partial_{\tilde t}\tilde v&=-\tilde x-\tilde\gamma \tilde
                               v+\tilde\xi(\tilde t),
  \label{eq:langevin_red}
\end{align}
with a Gaussian white noise $\tilde\xi(\tilde t)$ with correlations
$\langle\tilde\xi(\tilde t) \tilde\xi(\tilde t')\rangle=2\tilde\gamma\delta(\tilde
t-\tilde t')$ and $\tilde\gamma\equiv \gamma/(m\omega)$.
For the
remainder of this Supplemental Material, we drop the tilde.

The equilibrium
distribution in these reduced units is
\begin{equation}
  \label{eq:peq}
  \peq(x,v)=\frac{1}{2\pi}e^{-(x^2+v^2)/2}.
\end{equation}
We are interested in the fluctuations of the observable $\dot
N=|v|\delta(x)$, with mean
\begin{equation}
  \dN=\int dx\int dv\,
  \peq(x,v)\,\delta(x)|v|=1/\pi.
\end{equation}

The propagator for the
system, $p(x_1,v_1,t|x_0,v_0)$, gives the probability density to
find the system in state $(x_1,v_1)$ at time $t$
provided that it had been in state $(x_0,v_0)$ at time
$0$. With this propagator, the correlation function of the observable
$\dot N$
can be written as
\begin{align}
  \corrN{t}&=\int dx_0\int dv_0 \int dx_1\int dv_1\,
  \peq(x_0,v_0)\,\delta(x_0)|v_0|\,p(x_1,v_1,t|x_0,v_0)\,
  \delta(x_1)|v_1|\nonumber\\
  &=\int dv_0 \int dv_1\,|v_0v_1|\,
  \peq(0,v_0)\,p(0,v_1,t|0,v_0).
  \label{eq:corrint}
\end{align}

As the solution of a 2-dimensional
Ornstein-Uhlenbeck process, the propagator is Gaussian
\begin{equation}
  p(\vec z_1,t|\vec z_0)=\frac{1}{2\pi\sqrt{\det\vec{\sigma}}}e^{-(\vec{z}_1-\vec{\mu})^T\vec{\sigma}^{-1}(\vec{z}_1-\vec{\mu})/2},
\end{equation}
where we use a vectorial notation $\vec{z}\equiv(x,v)^T$.
The mean $\vec\mu$ and the covariance $\vec\sigma$ satisfy the
differential equations
\begin{align}
  \label{eq:mu}
  \dot{\vec{\mu}}&=-\vec{A}\vec{\mu}\\
  \label{eq:lyapunov}
  \dot{\vec{\sigma}}&=-\vec{A}\vec{\sigma}-\vec{\sigma}\vec{A}^T+\vec{D}
\end{align}
with the matrices
\begin{equation}
  \vec{A}\equiv\left(
  \begin{matrix}
    0&-1\\
    1&\gamma
  \end{matrix}\right),\qquad
  \vec{D}\equiv\left(
  \begin{matrix}
    0&0\\
    0&2\gamma
  \end{matrix}\right)
\end{equation}
and the initial conditions $\vec{\mu}(0)=\vec{z}_0$ and
$\vec{\sigma}(0)=\vec{0}$ (the $2\times 2$ matrix containing zeros).
For the evaluation of Eq.~\eqref{eq:corrint} we only need the initial
condition of the form $\vec{z}_0=(0,v_0)$, for which the solution of
Eq.~\eqref{eq:mu} yields $\vec{\mu}(t)=v_0\vec{\mu}_0(t)$ with
\begin{equation}
  \vec{\mu}_0(t)=\frac{1}{\bar\omega}e^{-\gamma t/2}\left(
  \begin{matrix}
    \sin(\bar\omega t)\\
    -(\gamma/2) \sin(\bar\omega t)+\bar\omega\cos(\bar\omega t)
  \end{matrix}\right)
\end{equation}
and the frequency of the damped oscillator
$\bar\omega\equiv\sqrt{1-\gamma^2/4}$. We consider only damping that
is below critical, i.e., $\gamma<2$. Solving the set of linear differential
equations for the coefficients of $\vec{\sigma}$ gives
\begin{equation}
  \vec{\sigma}(t)=
  \left(\begin{matrix}
    1&0\\
    0&1
  \end{matrix}\right)
+\frac{e^{-\gamma t}}{2\bar\omega}\left[
\left(\begin{matrix}
    -2&\gamma\\
    \gamma&-2
  \end{matrix}\right)
+
\left(\begin{matrix}
    \gamma/2&-1\\
    -1&\gamma/2
  \end{matrix}\right)
\gamma\cos(2\bar\omega t)
+
\left(\begin{matrix}
    -1&0\\
    0&1
  \end{matrix}\right)
\gamma\bar\omega\sin(2\bar\omega t)
  \right].
\end{equation}
Combining the Gaussian functions of the equilibrium distribution and
the propagator, the correlation function~\eqref{eq:corrint} can be
written as
\begin{equation}
  \corrN{t}=\frac{1}{4\pi^2\sqrt{\det\vec{\sigma}}}\int dv_0 \int dv_1\,|v_0v_1|\,
  \exp\left[-\frac{1}{2}  \left(\begin{matrix}
    v_0\\
    v_1
  \end{matrix}\right)\cdot\vec{\varSigma}\left(\begin{matrix}
    v_0\\
    v_1
  \end{matrix}\right)\right].
  \label{eq:corrint2}
\end{equation}
Therein, the matrix $\vec{\varSigma}$ is given as
\begin{equation}
  \vec{\varSigma}\equiv\left(
  \begin{matrix}
    1&0\\
    0&0
  \end{matrix}\right)+\vec{B}^T\vec{\sigma}^{-1}\vec{B}
\end{equation}
with
\begin{equation}
  \vec{B}\equiv\left(
  \begin{matrix}
    [\vec\mu_0(t)]_1&0\\
    [\vec\mu_0(t)]_2&1
  \end{matrix}\right).
\end{equation}
Finally, the Gaussian integral over $|v_0v_1|$ in Eq.~\eqref{eq:corrint2}
can be performed analytically, giving the result
\begin{equation}
  \corrN{t}=\frac{1}{\pi^2\sqrt{\det\vec\sigma}\det\vec\varSigma}\left(1+\frac{|\varSigma_{12}|}{\sqrt{\det\vec\varSigma}}\arctan \frac{|\varSigma_{12}|}{\sqrt{\det\vec\varSigma}}\right).
\end{equation}
The analytic expressions for the coefficients of the $2\times 2$
matrices can be recursively plugged into this result, however, the
resulting expression is lengthy and cannot be simplified.

For a straightforward numerical integration of the correlation function, it is
necessary that it does not show any divergence for $t\to 0^+$. Indeed,
expanding all the terms for small $t$ to leading order as $\det\vec\sigma\approx t^4/3$, $\det
\vec\varSigma\approx 3/t^2$, and $\varSigma_{12}\approx -1/t$, we get
the finite limit
\begin{equation}
  \lim_{t\to0^+}\corrN{t}=\gamma\frac{6\sqrt{3}+\pi}{18\pi^2}.
  \label{eq:smallt}
\end{equation}

\subsection{Timescale separation}

In our general minimal model for an escapement, we assume that the
distribution of the two possible states $y^-$ and $y^+$ of the counter
variable relaxes instantaneously to the equilibrium distribution $p_\pm$
determined by the rates $k^+$ and $k^-$. This assumption is justified if these
rates are large enough, such that the actual relaxation time of the counter
$1/k\equiv 1/(k^++k^-)$ is much smaller than any relevant timescale of the
oscillator process.

Yet, when we consider the underdamped harmonic oscillator, it may not be
obvious to identify its relevant timescales. Here, two subsequent ticks
(i.e., crossings of $x=0$) can be separated by a time
interval smaller than any finite $1/k$. The following analysis shows that such events
are sufficiently infrequent to not have a remaining contribution to the
statistics of the counter variable $y$ in the limit $k\to \infty$.

Assume there is a tick happening at time $t=0$, where the oscillator crosses
$x=0$ at speed $v_0$. Then, to generate the next tick, the oscillator needs to
be transported back to the position $x=0$, crossing it at speed $v_1$ with
sign opposite to $v_0$. As per the Langevin equation~\eqref{eq:langevin_red},
there is both a ballistic and a diffusive contribution to this transport. For
large $v_0$ and/or small $\gamma$, ballistic transport dominates, which causes
the next tick at the time $t\simeq\pi$. Hence $1/k\ll \pi$ is one condition on
the transition rates. Yet, for small $v_0$ and/or large $\gamma$, diffusive
transport provides a shortcut to having a tick much earlier.

\begin{figure}
  \centering
  \includegraphics{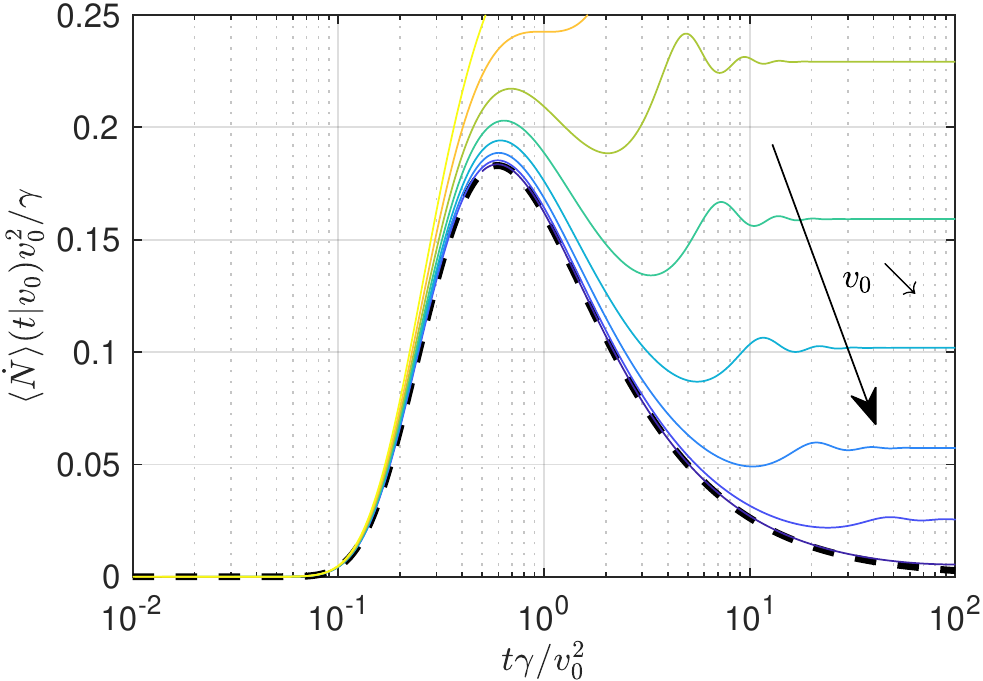}
  \caption{Rate of ticks in the time $t$ following a previous tick where the
    oscillator had the velocity $v_0$. Axes are scaled such that the scale
    invariant form (black dashed) that holds for $t\ll\pi$ becomes
    visible. Parameters: $\gamma=0.5$ and $v_0$ from $0.1$ (dark blue) to $0.8$ (yellow).}
  \label{fig:small_t}
\end{figure}

The rate of ticks at time $t>0$ conditioned on a tick at time $0$ at velocity
$v_0$ is given by
\begin{equation}
  \dN(t|v_0)\equiv\int d v_1\, |v_1|\,p(0,v_1,t|0,v_0),
  \label{eq:Ncond}
\end{equation}
with the Gaussian propagator introduced in the previous section. Results for
various values of $v_0$ are shown in Fig.~\ref{fig:small_t}. To assess the
behavior for small values of $t$, we expand every component of
$\vec{\sigma}(t)$ and $\vec{\mu}(t)$ to leading order in $t$. Performing the
Gaussian integral in Eq.~\eqref{eq:Ncond} then yields the scale invariant form
\begin{equation}
  \dN(t|v_0)=\frac{\gamma}{v_0^2}\phi(t\gamma/v_0^2)
  \label{eq:Ncondscale}
\end{equation}
with
\begin{equation}
  \phi(u)\equiv\frac{\sqrt{3}}{2\pi
    u}e^{-1/u}+\frac{\sqrt{3}}{4\sqrt{\pi}u^{3/2}}e^{-3/(4u)}\erf\left(\frac{1}{2u}\right).
  \label{eq:phidef}
\end{equation}
As shown in Fig.~\ref{fig:small_t}, this function has a distinct maximum at
$u\sim 1$, which corresponds to the re-crossings of $x=0$ caused by the
diffusive dynamics. For small $v_0^2/\gamma$, the corresponding maximum in
$\dN(t|v_0)$ can become indefinitely large. However, preceding this maximum
we see an onset time where re-crossings are exponentially suppressed, for
$t\gamma/v_0^2\ll 1$. This is the time it takes for the diffusive dynamics to
reverse the sign of $v$ in order to enable a repeated crossing of $x=0$. This
behavior is distinctive for underdamped dynamics, in contrast to overdamped
dynamics where every crossing of a point is immediately followed by an
infinite number of re-crossings.

If the counter equilibrates within the time where re-crossings are suppressed,
i.e., if $1/k\ll v_0^2/\gamma$, then the assumption that this equilibration
happens instantaneously is justified. Next, we calculate the fraction of ticks in the
stationary state where $|v_0|$
is below a certain small cutoff velocity $\hat v$,
\begin{equation}
  \frac{\langle\dot N|{-\hat v<v_0<\hat v}\rangle}{\langle\dot
    N\rangle}=\frac{1}{\langle\dot N\rangle}\int dx\int_{-\hat{v}}^{\hat{v}}
  dv\,|v|\,\delta(x)\,\peq(x,v)=1-e^{-\hat v^2/2}\approx \hat v^2/2.
\end{equation}
Thus, if $\gamma/k\ll 1$, or, more precisely, if we can find a $\hat v$ such
that $\gamma/k\ll \hat v^2\ll 1$, then the assumption of the separation of
timescales is justified for almost all ticks.

We now show explicitly that this separation of timescales leads indeed to the statistics
of the counter variable $y(t)$ derived in the main text. For this purpose we
consider the observables $N+(t)$ and $N^-(t)$, which count the number of
increments and decrements of $y(t)$ up to time $t$. Using the variable
$n(t)\in\{+,-\}$ that indicates whether the counter is in a state $y^+$ or
$y^-$, these observables can be defined analogously to $N(t)$ through
\begin{equation}
  N^\pm(t)=\int_0^td\tau |v(\tau)|\,\delta[x(\tau)]\,\delta_{n(\tau^-),\pm},
\end{equation}
where the Kronecker delta filters out ticks by the state of $n(t)$ ultimately
prior to the tick. Thus, we have $N(t)=N^+(t)+N^-(t)$ and
$y(t)=N^+(t)-N^-(t)$. The average of $y(t)$ can be written as
\begin{equation}
  \mean{y(t)}=t\int dv\,\peq(v)\,\left[\langle\dot N^+|v\rangle-\langle\dot
    N^-|v\rangle\right],
  \label{eq:meany}
\end{equation}
with steady state averages conditioned on the velocity $v$. Time reversal
symmetry in equilibrium requires that Eq.~\eqref{eq:Ncondscale} also gives the
rate of ticks at the time $t$ \textit{before} another tick, i.e.,
$\dN(t|v_0)=\dN(-t|v_0)$. Hence, we know that for $|v|>\hat v\gg\sqrt{\gamma/k}$
the variable $n(t)$ will have had time to equilibrate since the last tick,
such that $\langle\dot N^\pm|v\rangle=p^\pm\langle\dot N|v\rangle$.
We can therefore rewrite Eq.~\eqref{eq:meany} as
\begin{equation}
  \mean{y(t)}=t(p^+-p^-)\dN + t\int_{-\hat v}^{\hat v}
  dv\,\peq(v)\,\left[\langle\dot N^+|v\rangle-\langle\dot
    N^-|v\rangle-(p^+-p^-)\langle\dot N|v\rangle\right].
  \label{eq:meany2}
\end{equation}
For
$v\sim\sqrt{\gamma/k}$, the calculation of $\langle\dot N^\pm|v\rangle$ is
far less trivial (and probably not possible within the Gaussian
formalism). Yet, as we consider small $\hat v$, it only matters that the
integrand does not diverge. Indeed, since $0\leq\langle\dot N^\pm|v\rangle\leq\langle\dot N|v\rangle=\sqrt{2\pi}|v|$, we
find that the integral in Eq.~\eqref{eq:meany2} vanishes like $\hat
v^2$. The remaining first term is equal to the result obtained in the main
text based on the assumption of instantaneous equilibration.

For the evaluation of the dispersion of $y(t)$, we consider the integral
\begin{align}
  D_y=&\lim_{t\to\infty}\frac{1}{2t}\Var y(t)\nonumber\\&
  \approx\frac{1}{2}\dN+\int_{0^+}^{\hat t}
  d\tau\left[\smean{\dot N^+(0)\dot N^+(\tau)}-\smean{\dot N^+(0)\dot
  N^-(\tau)}-\smean{\dot N^-(0)\dot N^+(\tau)}+\smean{\dot N^-(0)\dot
  N^-(\tau)}-(\smean{\dot N^+}-\smean{\dot N^-})^2\right],
  \label{eq:Dyint}
\end{align}
where the upper limit $\hat t$ is chosen much larger than both the timescale
$t\sim\pi$ for oscillations and the timescale $t\sim1/\gamma$ for diffusion. On this
timescale, the dynamics of the system decorrelates, such that the
autocorrelation function of $\dot y(t)$ [the integrand of Eq.~\eqref{eq:Dyint}]
tends exponentially to zero. The choice of a large but finite $\hat t$ will be
convenient in the following analysis. The error incurred by this approximation
is exponentially small and therefore negligible.

If the equilibration of the counter after a tick was instantaneous, we would have
\begin{align}
  &\smean{\dot N^+(0)\dot N^+(\tau)}=p^+p^+\smean{\dot N(0)\dot N(\tau)},
  &\smean{\dot N^+(0)\dot N^-(\tau)}=p^+p^-\smean{\dot N(0)\dot N(\tau)},\nonumber\\
  &\smean{\dot N^-(0)\dot N^+(\tau)}=p^-p^+\smean{\dot N(0)\dot N(\tau)},
  &\smean{\dot N^-(0)\dot N^-(\tau)}=p^-p^-\smean{\dot N(0)\dot N(\tau)}.
    \label{eq:decorr}
\end{align}
Plugging this into Eq.~\eqref{eq:Dyint}, along with $\smean{\dot
  N^\pm}=p^\pm\dN$ then leads to the diffusion coefficient derived in the main
text. 

For large but finite $k$, deviations from Eq.~\eqref{eq:decorr} will surely occur for
times $\tau$ comparable to $1/k$. We therefore split the integral in
Eq.~\eqref{eq:Dyint} into $0<\tau<\hat v^2/\gamma$ and $\hat v^2/\gamma\leq
\tau<\hat t$. For the first part, we see in the following way that the
integrand cannot have any singularities:
Since $\dot N^\pm\geq 0$, all the four terms $\smean{\dot N^\pm(0)\dot
  N^\pm(\tau)}$ are bounded from below by zero. Moreover, they add up to
\begin{equation}
  \smean{\dot N^+(0)\dot N^+(\tau)}+\smean{\dot N^+(0)\dot N^-(\tau)}+\smean{\dot N^-(0)\dot N^+(\tau)}+\smean{\dot N^-(0)\dot N^-(\tau)}=\smean{\dot N(0)\dot N(\tau)}.
\end{equation}
Hence, each of the four terms is bounded from above by $\smean{\dot N(0)\dot
  N(\tau)}$, which we have already shown to be finite for small $\tau$ in
Eq.~\eqref{eq:smallt}. The part of the integral in Eq.~\eqref{eq:Dyint} with
$0<\tau<\hat v^2/\gamma$ therefore vanishes as $\hat v\to 0$.

For times $\tau\geq \hat v^2/\gamma$, deviations from Eq.~\eqref{eq:decorr}
are still possible. Even though there is enough time for $n(t)$ to equilibrate
between two ticks at times $0$ and $\tau$, further ticks may have occurred
briefly before these two points in time, bringing $n(t)$ out of
equilibrium. For a more detailed analysis, consider
\begin{equation}
  \int_{\hat v^2/\gamma}^{\hat t}
  d\tau\,\smean{\dot N^+(0)\dot N^+(\tau)}=
  \int_{\hat v^2/\gamma}^{\hat t}
  d\tau\int dv_0\int dv_1\,p(v_0,v_1,\tau)\,\smean{\dot N^+(0)\dot N^+(\tau)|v_0,v_1}
\end{equation}
[and analogously for the other three similar terms in Eq.~\eqref{eq:Dyint}]. Here we have
conditioned on the velocities $v_0$ and $v_1$ at times $0$ and $\tau$ and
averaged using their joint distribution $p(v_0,v_1,\tau)$. When both $v_0$
and $v_1$ are greater than $\hat v$, enough time will have passed since the
respective previous ticks to equilibrate the variable $n(t)$, such that we
obtain $\smean{\dot N^+(0)\dot N^+(\tau)|v_0,v_1}=p^+p^+\smean{\dot N(0)\dot
  N(\tau)|v_0,v_1}$. Writing
\begin{align}
  \int_{\hat v^2/\gamma}^{\hat t}
  d\tau\,\smean{\dot N^+(0)\dot N^+(\tau)}=&p^+p^+
  \int_{\hat v^2/\gamma}^{\hat t}
  d\tau\,\smean{\dot N(0)\dot N(\tau)}\nonumber\\ &+ \int_{\hat v^2/\gamma}^{\hat t}
  d\tau\int' dv_0\int' dv_1\,p(v_0,v_1,\tau)\,[\smean{\dot N^+(0)\dot
                                                    N^+(\tau)|v_0,v_1}-p^+p^+\smean{\dot
                                                    N(0)\dot
                                                    N(\tau)|v_0,v_1}],
 \label{eq:large_t}
\end{align}
where the integrals marked by $'$ run over the space where $v_0$ or $v_1$ are
smaller than $\hat v$, it now remains to be shown that the second line vanishes
as $\hat v\to 0$. Following the same logic as above, the terms $\smean{\dot
  N^\pm(0)\dot N^\pm(\tau)|v_0,v_1}$ are bounded from below by $0$ and from
above by $\smean{\dot
  N(0)\dot N(\tau)|v_0,v_1}$. Thus, it is sufficient to show that the integral
over the latter vanishes. Its part where $|v_0|<\hat v$ reads
\begin{equation}
  \int_{\hat v^2/\gamma}^{\hat t}
  d\tau\int_{-\hat v}^{\hat v} dv_0\int_{-\infty}^\infty dv_1\,p(v_0,v_1,\tau)\,\smean{\dot N(0)\dot N(\tau)|v_0,v_1}=  \int_{\hat v^2/\gamma}^{\hat t}
  d\tau\int_{-\hat v}^{\hat v} dv_0\,|v_0|\,\peq(0,v_0)\,\dN(\tau|v_0),
\end{equation}
where we identify the function from Eq.~\eqref{eq:Ncond}. The periodically
recurring ticks incurred by ballistic transport happen on average at a finite
rate $1/\pi$, their contribution to above integral must therefore vanish for
small $\hat v$. We therefore focus on the immediate re-crossings incurred by
diffusive transport, captured by the function $\phi(u)$ in
Eq.~\eqref{eq:Ncondscale},
\begin{align}
   \int_{\hat v^2/\gamma}^{\hat t}
  d\tau\int_{-\hat v}^{\hat v}
  dv_0\,|v_0|\,\peq(0,v_0)\,\frac{\gamma}{v_0^2}\phi(t\gamma/v_0^2)
  =  \int_{-\hat v}^{\hat v} dv_0\,|v_0|\,\peq(0,v_0)\, \int_{\hat v^2/v_0^2}^{\hat t\gamma/v_0^2}
  du\,\phi(u)\sim\hat v^2\ln\frac{\hat t\gamma}{\hat v^2}.
  \label{eq:v0int}
\end{align}
For the estimation of the value of the last integral we have used the fact
that we only need $\phi(u)$ for $u>1$, which is at first finite (see Fig.~\ref{fig:small_t}) and then, for
larger $u$, gets dominated by the first term in Eq.~\eqref{eq:phidef}
scaling like $u^{-1}$. As a result, we see that the integral vanishes as $\hat
v$ gets small.

For the remaining part of the integral in the second line of
Eq.~\eqref{eq:large_t} we calculate
\begin{align}
  &\int_{\hat v^2/\gamma}^{\hat t}
  d\tau\int_{|v_0|>\hat v}dv_0\int_{-\hat v}^{\hat v}
  dv_1\,p(v_0,v_1,\tau)\,\smean{\dot N(0)\dot N(\tau)|v_0,v_1}\nonumber\\
  &\qquad<  \int_{\hat v^2/\gamma}^{\hat t}
  d\tau\int_{-\infty}^\infty dv_0\int_{-\hat v}^{\hat v}
    dv_1\,p(v_0,v_1,\tau)\,\smean{\dot N(0)\dot N(\tau)|v_0,v_1}\nonumber\\
  &\qquad\qquad=\int_{\hat v^2/\gamma}^{\hat t}
  d\tau\int_{-\hat v}^{\hat v} dv_1\,|v_1|\,\peq(0,v_1)\,\dN(\tau|v_1)\sim\hat v^2\ln\frac{\hat t\gamma}{\hat v^2}.
\end{align}
In the second line, we have used the positivity of the integrand to extend the
integration space, and in the third line the symmetry of the correlation
function with respect to $\tau$ to arrive at the same integral as in
Eq.~\eqref{eq:v0int}. As a result, we see that the second line in
Eq.~\eqref{eq:large_t} vanishes altogether for small $\hat v$, such that the
simple expression for the dispersion of $y(t)$ from the main text is recovered.

\end{document}